# Pyrochlore NaYbO$_2$: A potential Quantum Spin Liquid Candidate

Chuanyan Fan, Tieyan Chang, Longlong Fan, Simon J. Teat, Feiyu Li, Xiaoran Feng, Chao Liu, Shilei Wang, Huifen Ren, Jiazheng Hao, Zhaohui Dong, Lunhua He, Shanpeng Wang, Chengwang Niu, Yu-Sheng Chen, Xutang Tao, and Junjie Zhang

**ABSTRACT:** The search for quantum spin liquids (QSL) and chemical doping in such materials to explore superconductivity have continuously attracted intense interest. Here, we report the discovery of a potential QSL candidate, pyrochlore-lattice β-NaYbO$_2$. Colorless and transparent NaYbO$_2$ single crystals, layered α-NaYbO$_2$ (~250 μm on edge) and octahedral β-NaYbO$_2$ (~50 μm on edge), were grown for the first time. Synchrotron X-ray single crystal diffraction unambiguously determined that the newfound β-NaYbO$_2$ belongs to the three-dimensional pyrochlore structure characterized by the $R\bar{3}m$ space group, corroborated by synchrotron X-ray and neutron powder diffraction and pair distribution function. Magnetic measurements revealed no long-range magnetic order or spin glass behavior down to 0.4 K with a low boundary spin frustration factor of 17.5, suggesting a potential QSL ground state. Under high magnetic fields, the potential QSL state was broken and spins order. Our findings reveal that NaYbO$_2$ is a fertile playground for studying novel quantum states.

## 1. INTRODUCTION

Quantum spin liquid (QSL) has attracted intense attention in the past several decades in two folds:[1-13] (i) QSL is relevant to high-temperature superconductivity.[4, 11, 14, 15] The nature of high-temperature superconductivity remains a central challenge in the fields of condensed matter physics, chemistry and materials science.[16-18] Nearly four decades ago, P. W. Anderson proposed QSL[13] to explain the mechanism of high temperature superconductivity in cuprates.[14] It was later found that the parent compounds of cuprates such as La$_2$CuO$_4$ are indeed Néel antiferromagnets.[19] However, the search for QSLs and chemical doping in such materials to look for high temperature superconductivity (To test Anderson's theory) have continuously attracted interest.[4, 15] (ii) QSL is an intriguing quantum state[13] and exhibits potential applications. QSL does not show any long-range magnetic order at low temperature even down to zero K but maintains highly entangled spins and strong fluctuations, making QSL very attractive for potential applications such as quantum computing and quantum information.[2-4, 9, 10] In real materials, QSL is likely to be found in low spin and geometrical frustrated systems including triangular, honeycomb, kagome, hyper-kagome, and pyrochlore lattice.[4] After long sought, a few materials such as κ-(ET)$_2$Cu$_2$(CN)$_3$,[20] herbertsmithite ZnCu$_3$(OH)$_6$Cl$_2$,[21] and α-RuCl$_3$[22] have been found to be promising candidate QSL materials; however, up to date, no QSL has been confirmed due to the lack of smoking-gun evidence.[9]

One main issue in candidate QSL materials is disorder.[4] It has been reported that there is 5~15% disordered Zn/Cu in herbertsmithite, making it challenging to explore intrinsic properties.[23] Experiments[24] and theoretical[25] arguments suggested that Ga/Mg disorder in YbMgGaO$_4$ is responsible for the disordered spin state and/or QSL stability. To find QSL systems with large exchange couplings and without disorder, the rare-earth chalcogenides ARX$_2$ (A= alkali or monovalent ions, R= rare earth, and X = O, S, Se) with a delafossite structure have been proposed,[26] and indeed, NaYbO$_2$,[27] NaYbSe$_2$,[28] and KYbSe$_2$[29] were reported to host QSL physics. In these materials, the Yb$^{3+}$ ions with effective $J_{eff}$ = 1/2 moments are supposed to form a perfect triangular lattice (space group $R\bar{3}m$) with alternating Yb layers and Na/K layers. Among them, delafossite NaYbO$_2$ (hereafter α-NaYbO$_2$) has attracted a lot of attention, and no signature of long-range magnetic order or spin glass appears down to 50 mK on polycrystalline powder samples.[26, 27, 30-33] In addition, it was reported that molar ratio of Na/Yb has a great impact on the physical properties of α-NaYbO$_2$.[30] A key feature of QSLs is that they support exotic spin excitations carrying fractional quantum numbers.[2, 3, 9, 10] However, measurements of these "fractionalized excitations" have been lacking due to the unavailability of α-NaYbO$_2$ single crystals.[26, 27, 30-33]

In this article, we report for the first time the growth of colorless and transparent NaYbO$_2$ single crystals with layered and octahedral morphologies using flux method. We performed synchrotron X-ray single crystal and powder diffraction, neutron powder diffraction and pair distribution function measurements to investigate their crystal structures. Although in-house X-ray powder diffraction patterns look identical, X-ray single crystal diffraction, in particular synchrotron X-ray single crystal diffraction, reveals that the layered crystals belong to the previously reported delafossite structure and the newfound octahedral crystals belong to pyrochlore structure with space group $R\bar{3}m$ (hereafter β-NaYbO$_2$) with *a* and *b* axes two times those of the delafossite structure. Magnetic susceptibility measurements on β-NaYbO$_2$ show no long-range magnetic order or spin glass behavior down to 0.4 K, indicating a potential QSL state.

Under high external magnetic fields, the potential QSL ground state in $\beta$-NaYbO$_2$ can be tuned to an ordered magnetic state, featuring a magnetization plateau at 1/3 saturation. Furthermore, we carried out theoretical calculations on the formation energy of $\alpha$- and $\beta$-NaYbO$_2$ to evaluate their thermodynamic stability, we found $\beta$-NaYbO$_2$ has lower formation energy, which is in good agreement with our crystal growth experiments.

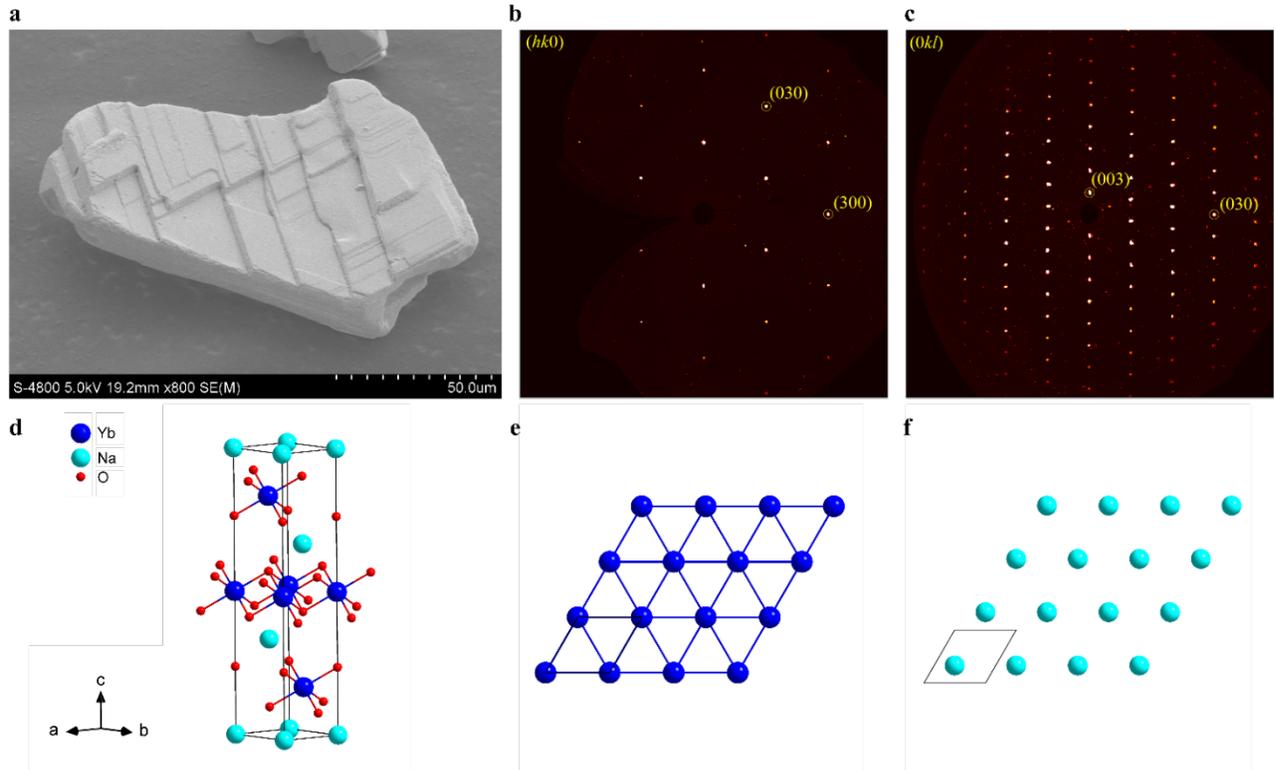

**Figure 1**. Crystal structure of $\alpha$-NaYbO$_2$. (a) SEM image of a typical layered $\alpha$-NaYbO$_2$ single crystal. (b) Reconstructed ($hk$0) plane from single crystal X-ray diffraction data. (c) Reconstructed (0$kl$) plane from single crystal X-ray diffraction data. (d) Three-dimensional crystal structure of $\alpha$-NaYbO$_2$. (e) The triangular arrangement of Yb atoms in the $ab$ plane. (f) Triangular layer formed by Na atoms in the $ab$ plane. Note the unit cell setting is $a = b = 3.3506(1)$ Å, $c = 16.5230(8)$ Å, $\alpha = \beta = 90°$ and $\gamma = 120°$.

## 2. RESULTS AND DISCUSSION

**2.1 Synthesis of polycrystalline powders.** NaYbO$_2$ polycrystalline powders were synthesized in the temperature range of 1000-1100 ºC with 10% excessive Na$_2$CO$_3$ using solid-state reaction according to the literature.[27, 33] All diffraction peaks match with those of $\alpha$-NaYbO$_2$ (PDF #01-085-7885), as shown in **Figure S1**. In order to determine suitable methods to grow single crystals, we heated the as-prepared powders to 1000 ºC to see whether NaYbO$_2$ is congruent melting or not. The powder diffraction pattern of the residual materials after heating show additional peaks from Yb$_2$O$_3$ (**Figure S2**). Thus, NaYbO$_2$ decomposes before it melts, indicating that melting methods (e.g., Czochralski and Bridgman) are not suitable and flux method is needed for single crystal growth.

**2.2 Single crystal growth using flux method.** The first and most important step for flux growth is finding a suitable flux, and self-flux is ideal because no other elements are introduced. Na$_2$O and Yb$_2$O$_3$ are considered, but their melting points are too high (1132 ºC for Na$_2$O and 2355 ºC for Yb$_2$O$_3$). We note that Na$_2$CO$_3$ has a relatively low melting point (851 ºC), and indeed, 10% excessive Na$_2$CO$_3$ helps stabilize pure NaYbO$_2$ powders, suggesting that Na$_2$CO$_3$ may be a good flux. Moreover, Na$_2$CO$_3$ has been used to prepare single crystals including BaFe$_{12}$O$_{19}$.[34] Initially, Na$_2$CO$_3$ and Yb$_2$O$_3$ were weighted in the mass ratio of 10:1, loaded in a Al$_2$O$_3$ crucible and heated to 1100 ºC, held for 24 h and then cooled to 900 ºC in 72 h, then the mixture was cooled from 900 ºC to 30 ºC in 10 hours, followed by furnace cooling to room temperature. Powders with tiny and shiny facets after washing with distilled water were obtained, and the powder diffraction pattern matches with $\alpha$-NaYbO$_2$ (PDF #01-085-7885), as shown in **Figure S1**. By changing concentration and temperature, single crystals of NaYbO$_2$ with plate and octahedral morphologies can be prepared. One major issue using Na$_2$CO$_3$ flux is low yield. To overcome this issue and grow large single crystals, we introduced NaOH (melting point 318 ºC) into Na$_2$CO$_3$ to reduce growth temperature.[35] By optimizing the ratio of Yb$_2$O$_3$: Na$_2$CO$_3$: NaOH and cooling rates, we have successfully grown NaYbO$_2$ single crystals with plate and octahedral morphologies (**Figure 1a, Figure 2a and Figure S3**). The morphology is mainly controlled by the growth temperature: the crystals prepared at ~1050 ºC have an octahedral morphology, while the crystals prepared at ~1100 ºC have a plate shape. These growth results are consistent with our theoretical calculations (see **2.8 Theoretical calculations**).

Inductively coupled plasma measurements revealed a molar ratio of Na:Yb to be 0.95:1 for plate crystals and 0.97:1 for octahedral crystals. Both are very close to 1:1. We also heated both plate and octahedral crystals to 1000 ºC and performed powder diffraction on residual materials (**Figure S2**). We found both of them decomposes, in good agreement with polycrystalline powders. In addition, both plate and octahedral crystals decompose after long-time grinding at ambient temperature or heating at 200 °C for 12 h in air (**Figure S4**). These results show that NaYbO$_2$ has relatively poor thermal stability, which suggest that it is challenging to prepare large size single crystals using seeded solution growth method.

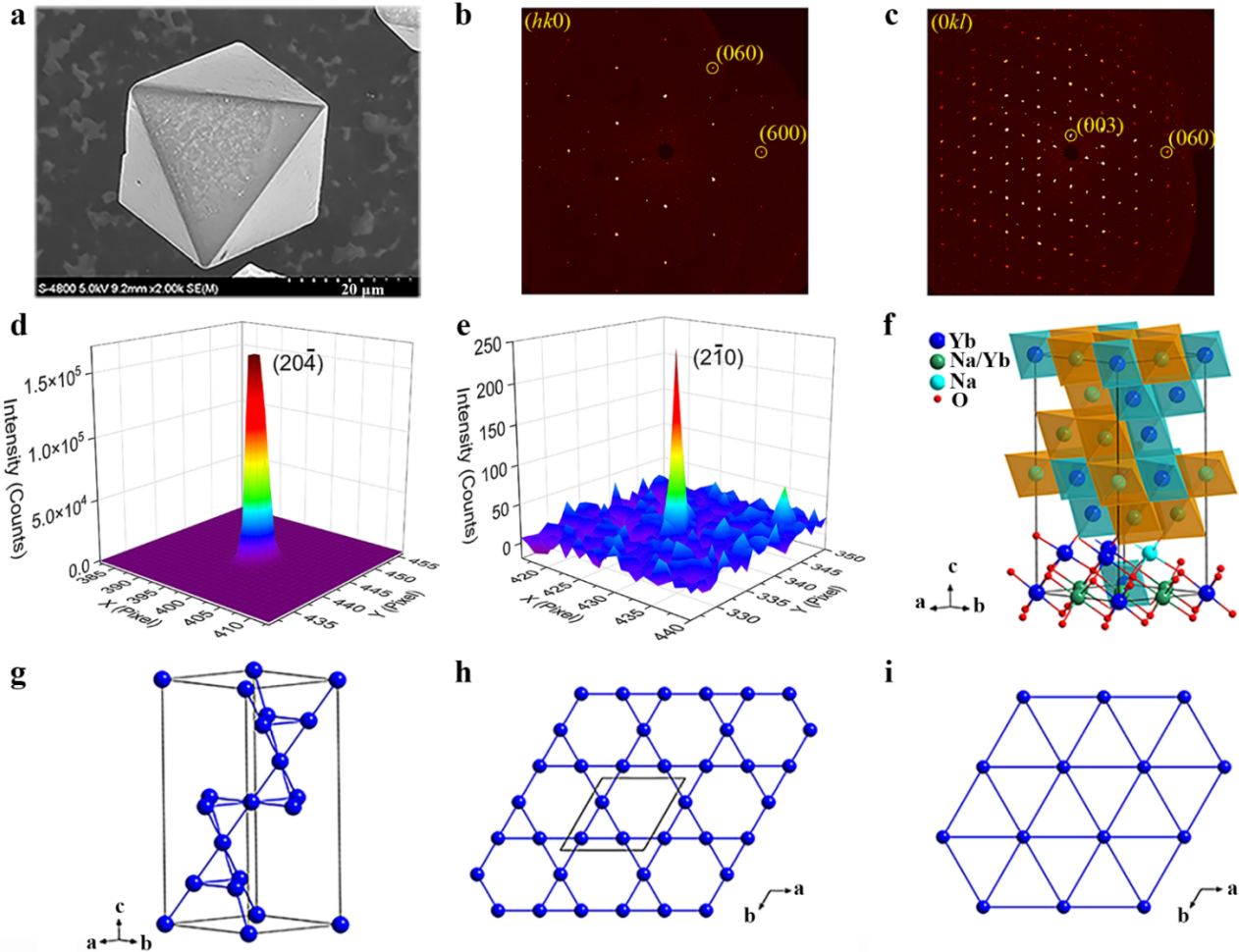

**Figure 2**. Crystal structure of β-NaYbO$_2$. (a) SEM image of a typical octahedral NaYbO$_2$ single crystal. (b) Reconstructed ($hk0$) plane from single crystal X-ray diffraction data. (c) Reconstructed ($0kl$) plane from single crystal X-ray diffraction data. (d) (2, 0, $\bar{4}$) peak. Note the peak is flat on the top due to saturation of detector. (e) (2, $\bar{1}$, 0) peak. (f) Three-dimensional crystal structure of NaYbO$_2$. (g) The pyrochlore arrangement of Yb atoms in the unit cell. (h) Kagome layer formed by Yb atoms at ($x, y, z$=1/6). (i) Triangular layer formed by Yb atoms at ($x, y, z$=0). Note the unit cell setting is $a = b$ = 6.7009(6) Å, $c$ = 16.402(2) Å, $\alpha=\beta$=90º and $\gamma$=120º.

The powder XRD diffraction patterns of the pulverized plate and octahedral crystals seem identical within the resolution of our in-house X-ray powder diffractometer (see **Note 1: More experiments and data analysis using laboratory X-rays, Figure S5 and Table S1**), and both match with those of polycrystalline powders synthesized at various temperatures and PDF #01-085-7885 (**Figure S1**). However, plate and octahedral single crystals belong to different structures: delafossite vs pyrochlore. We will unambiguously determine their structures in the following parts via single crystal X-ray diffraction, including synchrotron X-ray single crystal diffraction.

**2.3 Crystal structure of layered crystals (α-NaYbO$_2$).** We determine the crystal structure of as-grown plate single crystals using single crystal X-ray diffraction. As expected, plate NaYbO$_2$ crystallizes in the delafossite structure with space group $R\bar{3}m$ at 296 K ($a = b$ = 3.3506(1) Å, $c$ = 16.5230(8) Å). **Table S2** lists the crystallographic data at 296 K and **Table S3** presents the fractional atomic coordinates, isotropic thermal parameters, occupation and equivalent isotropic displacement parameters. **Figures 1b,c** show the reconstructued ($hk0$) and ($0kl$) planes from single crystal X-ray diffraction data consisting of 1457 frames collected at room temperature. **Figure 1d** shows the crystal structure of α-NaYbO$_2$ in the ball-and-stick style. The asymmetric unit contains one Yb, one Na and one O. All Yb and Na atoms are octahedrally coordinated by six O atoms with bond lengths of 2.507(2) Å for Na-O and 2.2554(19) Å for Yb-O octahedra. The Yb-O polyhedra form layers in the $ab$ plane (**Figure 1d**) and these layers alternately stack with Na-O layers along the $c$ axis (**Figure 1e,f**). In the Yb-O layer, Yb atoms form a perfect triangular lattice (**Figure 1e**) with a Yb-Yb distance of 3.3506(1) Å, and nearest Yb-Yb-Yb bond angles is 60°. Our structural model is corroborated by Rietveld refinement on synchrotron X-ray powder diffraction (**Figure S6** and **Table S4**), is also consistent with the literature.[27, 33]

**2.4 Crystal structure of octahedral crystals ($\beta$-NaYbO$_2$).** Surprisingly, in-house X-ray single crystal diffraction on octahedral crystals revealed a unit cell that is four times larger compared with the delafossite structure. Initially we suspected that the supercell was due to twinning and indeed twinning was observed in plate crystals (e.g., 89.77º between two domains, **Figure S7**). Careful examination on the reciprocal space excludes this possibility. **Figures 2b** and **Figure 2c** show reconstructed ($hk$0) and (0$kl$) planes using the setting of the rhombohedral lattice with $a=b\sim6.7$ Å and $c\sim16.4$ Å from single crystal X-ray diffraction data consisting of 2405 frames at room temperature. Reflections with $k$=odd, which are superlattice peaks with half integers using the delafosite lattice, are clearly seen in **Figure 2c**. To further verify our result, we selected the best single crystals for synchrotron X-ray single crystal diffraction at Advanced Light Source at Lawrence Berkeley National Laboratory. **Figure S8** shows the projections along $c$ axis and along $b$ axis in the reciprocal space (rhombohedral, $a=b\sim6.7$ Å and $c\sim16.4$ Å), and no twinning is found. **Figure 2d** shows the strongest Bragg peak (20$\bar{4}$), whose intensity (>1.64×10$^5$ counts) saturated the detector we used. In sharp contrast, the peak (2$\bar{1}$0) as shown in **Figure 2e**, which does not exist in the delafossite structure, is at least three orders of magnitude weaker. The high quality synchrotron X-ray single crystal diffraction data unambiguously determined the pyrochlore-lattice structure of the as-grown NaYbO$_2$ single crystals with octahedral morphology.

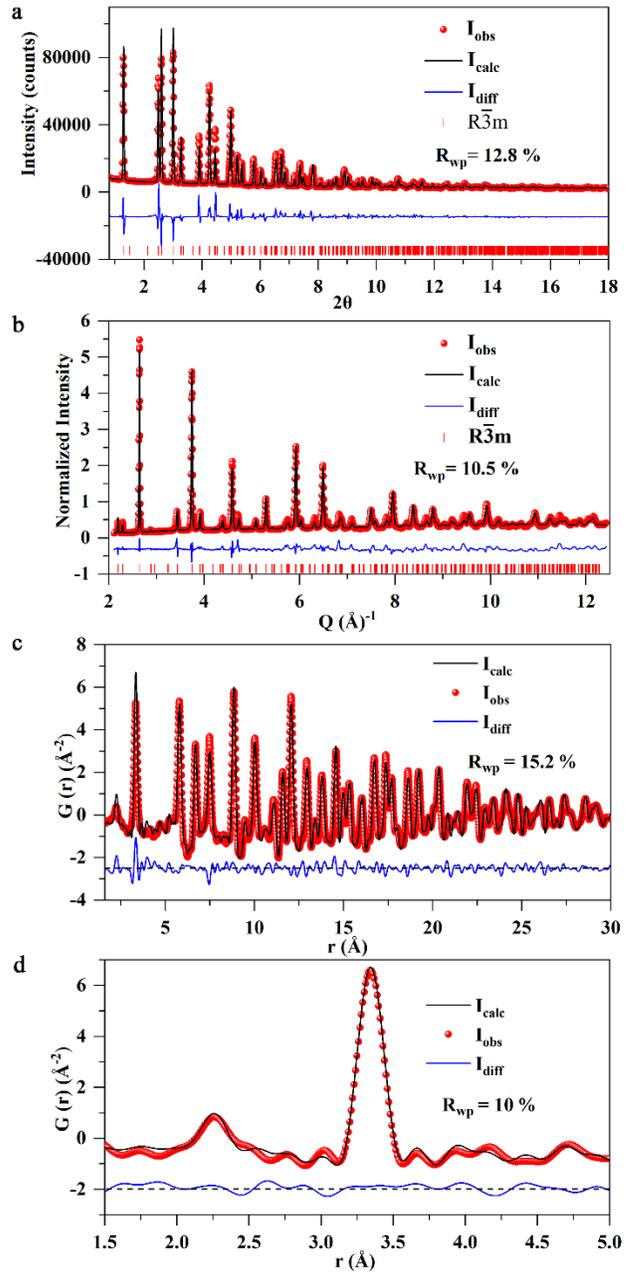

**Figure 3**. Rietveld refinements on pulverized $\beta$-NaYbO$_2$ single crystals. (a) Synchrotron X-ray powder diffraction pattern collected at room temperature and Rietveld refinement using the single crystal structural model. (b) Neutron powder diffraction pattern at room temperature and Rietveld refinement. (c) and (d) Pair distribution function (PDF) at room temperature and Rietveld refinement.

The structure of $\beta$-NaYbO$_2$ belongs to the 3D pyrochlore structure characterized by the $R\bar{3}m$ space group (No. 166). **Table S2** lists the crystallographic data at 200 K and **Table S3** presents the fractional atomic coordinates, isotropic thermal parameters and occupation. There are two Yb atoms, two Na atoms and two O atoms in the asymmetric unit. One of the Na positions (Na1) is disordered with Yb with occupation of 0.072(8) for Yb. All Yb and Na atoms are modeled using anisotropic displacement parameters. The Yb atoms are surrounded by six O atoms with bond distances of 2.257(9)-2.271(8) Å. The Na atoms are coordinated by six O atoms with bond lengths of 2.475(10)-2.489(9) Å. **Figure 2f** shows 3D crystal structure of β-NaYbO₂ in the polyhedral style. Notably in the *ab* plane, Na-O and Yb-O polyhedra are alternatively arranged along *a* or *b* axis rather than Na layers alternating with Yb layers found in the delafossite structure (**Figure 1**). **Figure 2g** shows the distribution of Yb atoms in the unit cell. Yb atoms (Yb2) with $z$=1/6, 3/6 and 5/6 form a kagome structure with Yb-Yb nearest distance of 3.3505(4) Å (**Figure 2h** and **Figure S9**), while Yb atoms (Yb1) with $z$=0, 2/6 and 4/6 form a triangular structure with Yb-Yb nearest distance of 6.7009(6) Å, and nearest Yb-Yb-Yb bond angles is 60° (**Figure 2i** and **Figure S9**). The pyrochlore network of Yb and Yb-Yb distance of β-NaYbO₂ is quite similar to the quantum material Yb₂Ti₂O₇,[36-39] suggesting that β-NaYbO₂ may host exotic physical properties. The 3D crystal structure is consistent with the octahedral morphology of the as-grown single crystals. A similar 3D pyrochlore structure was previously reported in LiYbSe₂ which crystallizes in the $Fd\bar{3}m$ space group.[40] The symmetry of our β-NaYbO₂ is lower than $Fd\bar{3}m$, because the powder diffraction pattern shows clear peak splitting at around 54º (**Figure S1**).

We calculated the X-ray powder diffraction patterns of both α- and β-NaYbO₂ using our single crystal models using λ=1.54059 Å in the 2θ range of 10-120°. As shown in **Figure S10**, more diffraction peaks are observed in β-NaYbO₂, compared with that of α-NaYbO₂. However, the intensities of these extra peaks are too weak to be observed using the in-house X-ray powder diffraction. For example, the peaks located at 2θ=18.7° and 26.6° are the strongest two superlattice peaks, but their intensities are at least three orders in magnitude smaller than that of the strongest Bragg peak (**Table S5**). This explains why we do not observe any difference in powder diffraction patterns of α-NaYbO₂ and β-NaYbO₂ using in-house X-rays as shown in **Figure S1**, and β-NaYbO₂ is a "hidden" polymorph of NaYbO₂.

**2.5 Rietveld refinements on powder data of β-NaYbO₂.** Rietveld refinements on synchrotron X-ray powder diffraction data, neutron powder diffraction data and pair distribution function data corroborate our β-NaYbO₂ single crystal structural model. **Figure 3a** shows the synchrotron X-ray powder diffraction data collected at Beijing Synchrotron Radiation Facility. Rietveld refinement converged to $R_{wp}$=12.8 % and GOF=9.7. The obtained atomic positions and displacement parameters are listed in **Table S4**, which are consistent with single crystal data. **Figure 3b** presents the neutron diffraction data collected at General Purpose Powder Diffractometer (GPPD). The Rietveld refinement converged to $R_{wp}$=10.5 % and GOF=6.47. To study local structure, we carried out pair distribution function measurements on β-NaYbO₂ at room temperature using synchrotron X-rays. As can be seen from **Figure 3c** and **Figure 3d**, the β-NaYbO₂ structure model is satisfactory in fitting the data in the range of 1.5-30 Å and 1.5-5.0 Å. Moreover, we collected PDF data of both α- and β-NaYbO₂ at Shanghai Synchrotron Radiation Facility for comparison (**Figure S11**). Obvious difference in G(*r*) at around 3.35 Å is observed, confirming the existence of two different phases. This result also indicates the probability of nearest Yb-Yb, Yb-Na and Na-Na is different between the two polymorphs.

**2.6 Optical properties.** We measured the ultraviolet-visible (UV-Vis) spectra, Infrared spectra, and Raman spectra of pulverized single crystals of α-NaYbO₂ and β-NaYbO₂. **Figure S12** shows the UV-Vis absorption spectra of polycrystalline powder synthesized from solid state reactions, pulverized α-NaYbO₂ and β-NaYbO₂ crystals. **Table S6** lists the calculated bandgap from UV-Vis absorption spectra (see **Figure S13** for calculations of bandgap). Within the resolution of our measurements, α-NaYbO₂ and β-NaYbO₂ show similar bandgaps. Such a wide bandgap is consistent with the color of the as-grown single crystals. The Raman shift is related to the vibration and rotational energy levels of the substance molecules. As shown in **Figure S14**, the peak positions of the Raman spectra of α-NaYbO₂ and β-NaYbO₂ are identical. In addition, we also performed FTIR measurements on the pulverized α-NaYbO₂ and β-NaYbO₂ crystals. The absorption peaks near 1500 cm⁻¹ and 1380 cm⁻¹ belong to NaYbO₂. Thus, optical properties including bandgap, IR and Raman spectra are indistinguishable for α-NaYbO₂ and β-NaYbO₂.

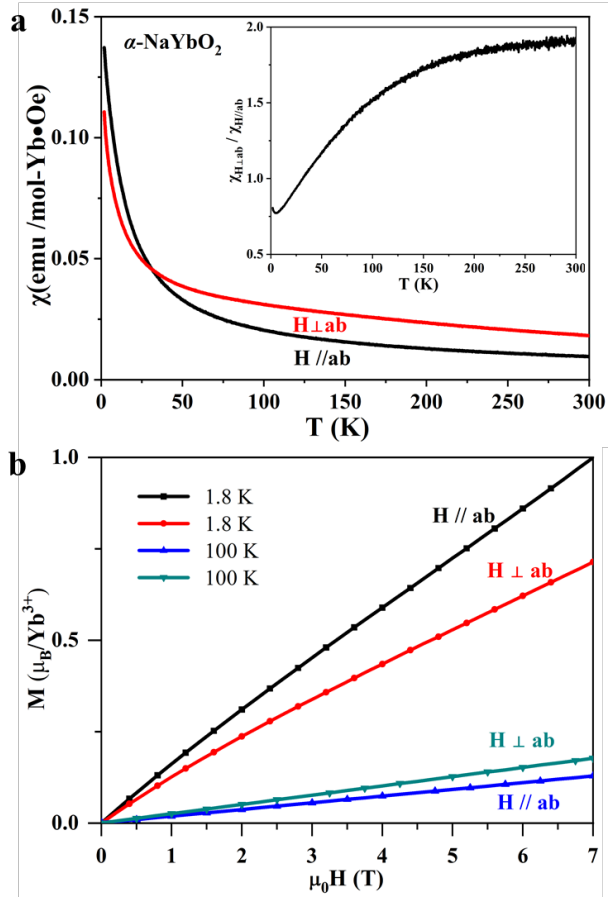

**Figure 4.** Magnetic susceptibility and magnetization of co-aligned α-NaYbO₂ single crystals. (a) Magnetic susceptibility measured under 0.01 T using MPMS3 VSM. The inset shows the ratio of $\chi_{H\perp ab}/\chi_{H//ab}$. (b) Magnetization as a function of magnetic field at T=1.8 and 100 K.

**2.7 Magnetic properties.**

**2.7.1 α-NaYbO₂.** The absence of long-range magnetic order and spin glass behavior under low field is found down to 0.4 K on pulverized single crystals of α-NaYbO₂ (**Figure S15**), consistent with previous reports on powders.[26, 27, 31, 32] Due to relatively small size of single crystals, it's difficult to measure direction dependent magnetic properties on a single crystal. However, we are able to co-align single crystals of ~0.7 mg to measure in-plane and out-of-plane magnetic properties. **Figure 4a** shows the in-plane and out-of-plane magnetic susceptibility under an external magnetic field of 100 Oe in the temperature range of 1.8-300 K (see raw data in **Figure S16**). Curie-Weiss fit were performed (**Figure S17**). The ratio of $\chi_{H\perp}$

$_{ab}/\chi_{H//ab}$, quantifying anisotropy of magnetic susceptibility, decreases from 1.9 at room temperature to 0.8 at 1.8 K with decreasing of temperature. Interestingly, there is a crossover at temperature of ~32 K. The in-plane and out-of-plane magnetization as a function of field is presented in **Figure 4b** and **Figure S16**. As reported by Bordelon et al.,[27] the quantum disordered ground state in $\alpha$-NaYbO$_2$ can be tuned into an up-up-down antiferromagnetic ordered regime in magnetic fields, stabilizing a quantized magnetization plateau at 1/3 saturation. To check if our single crystals show such a feature, we measured magnetic susceptibility between 0.4 and 1.8 K under high fields and magnetization at 0.4 K as a function of field, as shown in **Figure S18**. $\alpha$-NaYbO$_2$ is found to maintain a quantum disordered state under 4 T but enter into an ordered magnetic state under 5 T. Different behaviors are observed between in-plane and out-of-plane magnetic susceptibility under 4 and 5 T. The location of the phase boundary between quantum disordered and ordered states is slightly different from Bordelon et al.,[27] which is likely due to field polarized of the crystals and the known $g$-tensor anisotropy of the material.[27, 32] As expected, we see a field-induced transition in magnetization. It's worth noting that the ~1/3 plateau is clearly seen in the in-plane magnetization but weakly in the out-of-plane magnetization, implying that the spin direction in the ordered state is mainly out of the $ab$ plane. Our result is consistent with Bordelon et al.[27]

**2.7.2 $\beta$-NaYbO$_2$.** The $\beta$-NaYbO$_2$ single crystals are too small and it is very challenging to co-align them for direction dependent measurements, thus we pulverized single crystals to make dense pellets for measurements. **Figure 5a** shows magnetic susceptibility data in the temperature range of 1.8-300 K under a magnetic field of 20 Oe. No anomaly or no bifurcation between the ZFC and FC curves, typically found in spin-glass systems, is observed. Below 50 K, the Van Vleck contribution to the susceptibility is negligible. The ZFC-W data between 5 and 20 K are fit to Curie-Weiss Law (**Figure 5a** inset) using $\chi(T) = \chi_0 + C/(T - \theta_{CW})$, where $\chi_0$ is a temperature independent term, $C$ is the Curie constant, and $\theta_{CW}$ is the Weiss temperature. $\chi_0 = 3.2\times10^{-3}$ emu Oe$^{-1}$ mol$^{-1}$, $C$ = 0.98 emu K$^{-1}$ mol$^{-1}$ and $\theta_{CW}$ = -7.0 K are obtained. The negative $\theta_{CW}$ indicates antiferromagnetic interactions. The obtained moment $\mu_{eff}$ = 2.8 $\mu_B$ is reminiscent of an effective $J$ =1/2 state. The data in the range of 200-300 K (**Figure S19**) are also fit using the Curie-Weiss law, which results in $\chi_0 = 7.8\times10^{-5}$ emu Oe$^{-1}$ mol$^{-1}$, $C$ = 2.72 emu K$^{-1}$ mol$^{-1}$ and $\theta_{CW}$ = -99.9 K. An effective moment of $\mu_{eff}$ = 4.66 $\mu_B$ is obtained, consistent with the theoretically predicted 4.54 $\mu_B$ for Yb$^{3+}$ ions.[41] **Figure 5b** show the field-dependent magnetization at various temperatures between 1.8 and 300 K. At high temperatures, linear behavior is observed. The field-dependent magnetization shows a nonlinear behavior at 1.8 K without saturation at $\mu_0$H = 7 T. The observed maximum magnetization M$_s$ = 0.94 $\mu_B$/Yb$^{3+}$ at 7 T is much smaller than the expected saturated moment for this system (~1.5 $\mu_B$ per Yb).[27]

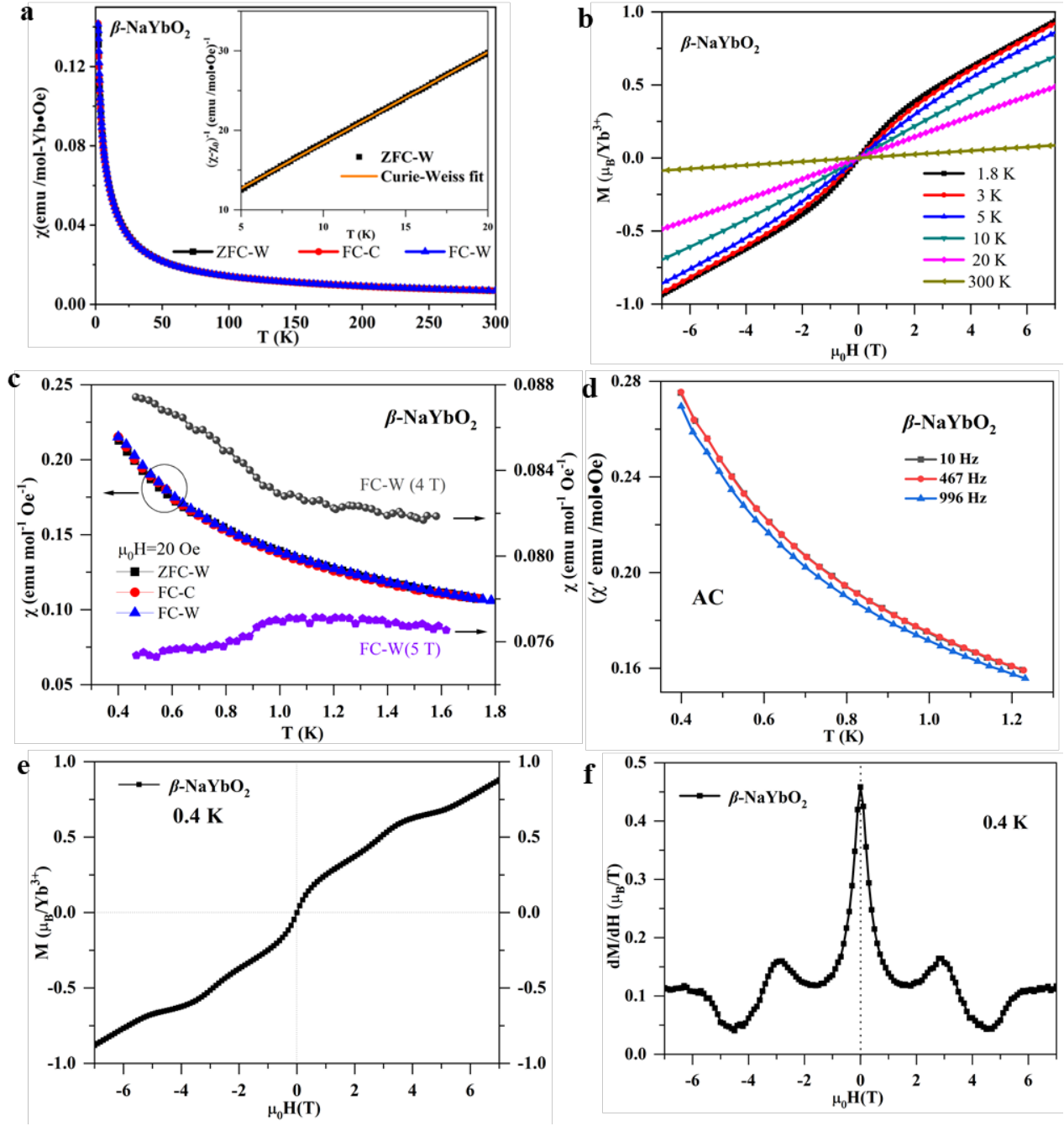

**Figure 5.** Magnetic susceptibility and magnetization of pulverized single crystals of $\beta$-NaYbO$_2$. (a) Magnetic susceptibility under 20 Oe in the range of 1.8-300 K. The insets show the Curie-Weiss fit in the temperature range of 5-20 K. (b) Magnetization as a function of magnetic field at 1.8, 3, 5, 10, 20 and 300 K. (c) Magnetic susceptibility between 0.4 and 1.8 K under 20 Oe, 4 T and 5 T. (d) Temperature and frequency dependence of a.c. magnetic susceptibility $\chi'(T)$ from 0.4 to 1.25 K. (e) Magnetization as a function of magnetic field at 0.4 K. (f) The first derivatives of magnetization data shown in (e).

Now we turn to low temperature data between 0.4 and 1.8 K (no phase degradation occurs, see **Figure S20** for PXRD before and after measurements). As shown in **Figure 5c**, the magnetic susceptibility under a magnetic field of 20 Oe increases monotonically with decreasing temperature and no anomaly is found. **Figure 5d** shows the AC magnetic susceptibility data. No signatures of spin freezing or frequency dependence are observed. Ramirez has provided a measure for spin frustration by defining $f = |\theta_{CW}|/T_c$.[42] In our case, we take the $\theta_{CW}$ from the CW fit between 5 and 20 K and $T_c$ = 0.4 K to estimate the lower boundary for $f$. We obtain a minimum $f$ = 17.5, which is already much larger than 10, signifying a strong frustration effect. Our results indicate that $\beta$-NaYbO$_2$ exhibit a potential QSL ground state. Ultralow temperature magnetic measurements and fractionalized excitations using inelastic neutron scattering are demanded to further support this deduction.

Like $\alpha$-NaYbO$_2$,[27] the magnetic ground state in $\beta$-NaYbO$_2$ can be tuned by external magnetic fields to an ordered state, as shown in magnetic susceptibility under 4 and 5 T (**Figure 5c**).

The magnetization data as a function of field at 0.4 K is shown in **Figure 5e**. Such a magnetization plateau, indicating a field-induced quantum state, has been reported in other Yb-based chalcogenide compounds.[43] The first derivative of MH is shown in **Figure 5f**. Two peaks at 2.9 and 5.6 T are clearly seen. We have analyzed the isothermal magnetization data in **Figure 5e** with the linear part subtracted. Similar data analysis has been reported previously.[44-47] Specifically, we first fit the linear parts using the linear fit M = $a\mu_0 H + b$, and then subtracted from the original data. **Figures S21** shows the net magnetization with linear components subtracted. A plateau, M = 0.45 $\mu_B$/Yb between 4.3 and 4.8 T for $\beta$-NaYbO$_2$, is clearly seen at 1/3 saturation (0.45/1.36=0.33). Similar phenomena has been reported in other triangular-lattice antiferromagnets, including Cs$_2$CuBr$_4$[48, 49] and Ba$_3$CoSb$_2$O$_9$.[50] Neutron diffraction would be highly desirable to understand the exact nature of this ordered phase in the pyrochlore polymorph of NaYbO$_2$.

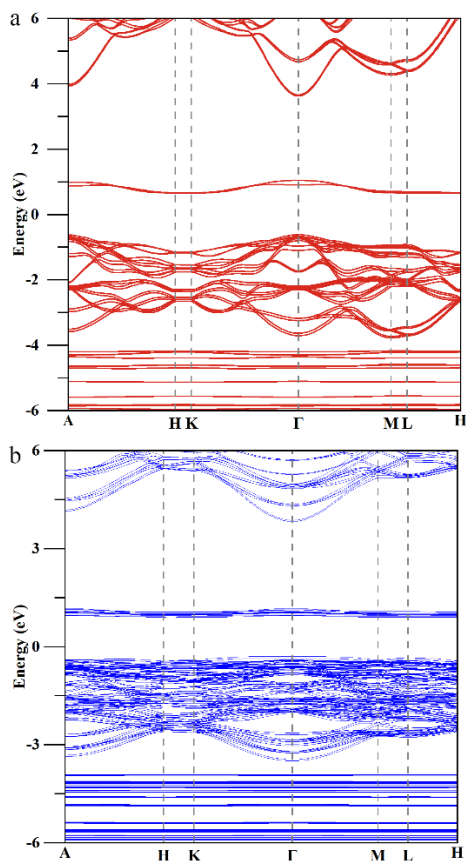

**Figure 6**. Calculated band structures of (a) $\alpha$-NaYbO$_2$ and (b) $\beta$-NaYbO$_2$ with spin-orbit coupling.

**2.8 Theoretical calculations.** To evaluate the stability of Na$_m$Yb$_m$O$_{2m}$ (m=3 for $\alpha$-NaYbO$_2$ and m=12 for $\beta$-NaYbO$_2$), formation energy is calculated by $E_f = (E_{\text{catal}} - m\mu_{Na} - m\mu_{Yb} - 2m\mu_O)/4m$, where $E_{\text{catal}}$ is the total energy of the compound, and $\mu$ represents the energy of a single Na, Yb, or O atom in its most stable bulk phase. The optimized lattice constants are $a$= 3.347 Å and $c$= 16.550 Å for $\alpha$-NaYbO$_2$, and $a$= 6.759 Å and $c$= 16.554 Å for $\beta$-NaYbO$_2$. The obtained formation energies are -2.30 eV/atom for $\alpha$-NaYbO$_2$ and -2.42 eV/atom for $\beta$-NaYbO$_2$, indicating their high thermodynamic stability. More importantly, the lower formation energy of $\beta$-NaYbO$_2$ suggests that it is easier to synthesize this polymorph, which is consistent with the lower temperature used for single crystal growth (see **2.2 Single crystal growth using flux method**).

**Figure 6** presents the band structures of $\alpha$-NaYbO$_2$ and $\beta$-NaYbO$_2$ with spin-orbit coupling (SOC), calculated using the ferromagnetic (FM) spin configuration, i.e., with the spins of Yb$^{3+}$ ions forced to align along the $c$-axis. GGA+U calculations reveal that both $\alpha$-NaYbO$_2$ and $\beta$-NaYbO$_2$ are insulators, and the band gaps for $\alpha$-NaYbO$_2$ and $\beta$-NaYbO$_2$ are 1.27 eV and 1.22 eV, respectively. The calculated band gaps are almost the same for both $\alpha$-NaYbO$_2$ and $\beta$-NaYbO$_2$, consistent with experiments; however, they are significantly lower than the experimental values (~4.8 eV). This discrepancy indicates that the GGA+U method, due to its oversimplified mean-field treatment of the on-site Coulomb U, fails to accurately describe the strongly correlated materials $\alpha$-NaYbO$_2$ and $\beta$-NaYbO$_2$. Similar results have been reported previously in $\alpha$-NaYbO$_2$.[51] Accurately calculating the electronic structures of $\alpha$-NaYbO$_2$ and $\beta$-NaYbO$_2$ requires advanced many-body techniques, such as Dynamical Mean Field Theory (DMFT), to address strong correlation effects.

## 3. CONCLUSION

We have successfully grown layered $\alpha$-NaYbO$_2$ and octahedral $\beta$-NaYbO$_2$ single crystals using flux method. For the previously reported $\alpha$-NaYbO$_2$ polymorph, single crystals were grown for the first time, and single crystal X-ray diffraction confirms the layered delafossite structure. For the newly discovered $\beta$-NaYbO$_2$ polymorph, synchrotron X-ray single crystal diffraction unambiguously determined that $\beta$-NaYbO$_2$ belongs to the 3D pyrochlore structure characterized by $R\bar{3}m$ with lattice parameters of $a = b$ = 6.7009 (6) Å, $c$ = 16.402(2) Å, corroborated by synchrotron X-ray and neutron powder diffraction and pair distribution function analysis. Magnetic susceptibility measurements revealed strong AFM interactions and no long-range magnetic order or spin glass behavior down to 0.4 K. The lower boundary of spin frustration factor was estimated to be 17.5, signifying a strong effect in spin frustration. At low magnetic fields, pyrochlore $\beta$-NaYbO$_2$ is a potential QSL candidate, while at high magnetic fields, the material undergoes quantum phase transitions to become magnetic ordered. Our findings provide a new QSL candidate in the rare-earth chalcogenides family that host pyrochlore-type structure in addition to the previously reported delafossite structure.

## ASSOCIATED CONTENT

**Supporting Information**. This material is available free of charge via the Internet at http://pubs.acs.org.

Experimental details for solid state reaction, single crystal growth, SEM, ICP-MS, UV-Vis-NIR, FTIR, powder X-ray diffraction, single crystal X-ray diffraction, pair distribution function, neutron powder diffraction, magnetic susceptibility and magnetization measurements, and theoretical calculations; Note 1 More experiments and data analysis on two polymorphs using laboratory x-rays; Photos of as-grown single crystals; figures of powder diffraction and Rietveld refinement/peak fitting, reciprocal space of single-crystal diffraction data, structural model, PDF data, UV-Vis absorption spectra, bandgap, Raman and IR spectra, Curie-Weiss fit, and magnetization as a function of field and magnetization plateaues; and tables of crystallographic data, fractional atomic positions and equivalent isotropic displacement parameters, calculated powder diffraction intensities, and peak position, intensity and width of the four strongest peaks and peak splitting at ~54º for $\alpha$- and and $\beta$-NaYbO$_2$ using laboratory x-ray powder diffraciton (PDF)

## AUTHOR INFORMATION

### Corresponding Author

***Junjie Zhang -** Institute of Crystal Materials, State Key Laboratory of Crystal Materials, Shandong University, Jinan 250100, Shandong, China; orcid.org/0000-0002-5561-1330; E-mail: junjie@sdu.edu.cn

### Authors

**Chuanyan Fan -** Institute of Crystal Materials, State Key Laboratory of Crystal Materials, Shandong University, Jinan 250100, Shandong, China

**Tieyan Chang -** NSF's ChemMatCARS, University of Chi cago, Lemont, Il 60439, United States

**Longlong Fan -** Institute of High Energy Physics, Chinese Academy of Sciences, Beijing 100049, Beijing, China

**Simon J. Teat -** Diffraction & Imaging, Photon Science Opera tions, Advanced Light Source, Lawrence Berkeley National Laboratory, CA 94720, United States

**Feiyu Li -** Institute of Crystal Materials, State Key Laboratory of Crystal Materials, Shandong University, Jinan 250100, Shandong, China

**Xiaoran Feng -** School of Physics, Shandong University, Jinan 250100, Shandong, China

**Chao Liu -** Institute of Crystal Materials, State Key Laboratory of Crystal Materials, Shandong University, Jinan 250100, Shandong, China

**Shilei Wang -** Institute of Crystal Materials, State Key Laboratory of Crystal Materials, Shandong University, Jinan 250100, Shandong, China

**Huifen Ren -** Huairou Division, Institute of Physics, Chinese Academy of Sciences, Beijing 101407, Beijing, China

**Jiazheng Hao -** Spallation Neutron Source Science Center, Dongguan 523803, China

**Zhaohui Dong –** Shanghai Synchrotron Radiation Facility (SSRF), Shanghai Advanced Research Institute Chinese Academy of Science, Shanghai 201204, China

**Lunhua He -** Spallation Neutron Source Science Center, Dongguan 523803, China

**Shanpeng Wang -** Institute of Crystal Materials, State Key Laboratory of Crystal Materials, Shandong University, Jinan 250100, Shandong, China

**Chengwang Niu -** School of Physics, Shandong University, Jinan 250100, Shandong, China

**Yu-Sheng Chen -** NSF's ChemMatCARS, University of Chicago, Lemont, Il 60439, United States

**Xutang Tao -** Institute of Crystal Materials, State Key Laboratory of Crystal Materials, Shandong University, Jinan 250100, Shandong, China

### Author Contributions

The manuscript was written through contributions of all authors.

### Notes

The authors declare no competing financial interest.

## ACKNOWLEDGMENT

J.Z. thanks Prof. Yang Ren from University of City Hong Kong for his help with synchrotron X-ray powder diffraction. J.Z. thanks Prof. Jian Zhang for his help with in-house single crystal X-ray diffraction. Work at Shandong University was supported by the National Natural Science Foundation of China (Grant No. 12374457 and 12074219), the Taishan Scholars Project of Shandong Province (tsqn201909031), and the QiLu Young Scholars Program of Shandong University. Synchrotron X-ray powder diffraction and pair distribution function measurements were performed at Beijing Synchrotron Radiation Facility (BSRF) at Beijing Institute of High Energy Physics, Chinese Academy of Sciences. The Advanced Light Source was supported by the Director, Office of Science, Office of Basic Energy Sciences, of the U.S. Department of Energy under Contract No. DE-AC02-05CH11231. Measurements of magnetic properties below 2 K were carried out at the Synergetic Extreme Condition User Facility (SECUF). Neutron powder diffraction experiments were performed at the General Purpose Powder Diffractometer (GPPD) at China Spallation Neutron Source (CSNS) under user program (Proposal no. P1822113000002). The work was partially performed at BL12SW at Shanghai Synchrotron Radiation Facility (SSRF).

For Table of Contents Use Only

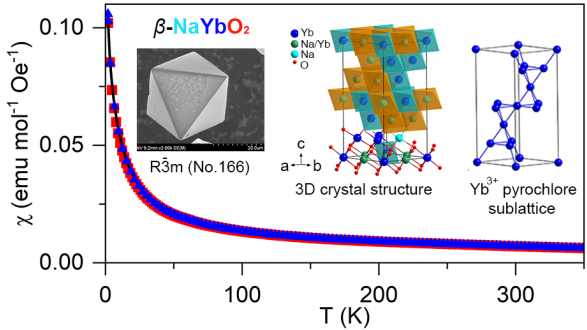